\documentclass[prl,floatfix,twocolumn,aps,showpacs,superscriptaddress]{revtex4-1}
\usepackage{graphicx}
\usepackage{amsmath}
\usepackage{amssymb}
\usepackage{natbib}
\usepackage{color}

\begin{document}
	
	\newcommand{\beq}{\begin{equation}}
	\newcommand{\eeq}{\end{equation}}
	
	\title{Excess Floppy Modes {and Multi-Branched Mechanisms} \\in Metamaterials with Symmetries}
	
	\author{Luuk A. Lubbers}
	\affiliation{Huygens-Kamerlingh Onnes Lab, Universiteit Leiden, P.O.~Box~9504, NL-2300 RA Leiden, The Netherlands}
	\affiliation{AMOLF, Science Park 104, 1098 XG Amsterdam, The Netherlands}
	\author{Martin van Hecke}
	\affiliation{Huygens-Kamerlingh Onnes Lab, Universiteit Leiden, P.O.~Box~9504, NL-2300 RA Leiden, The Netherlands}
	\affiliation{AMOLF, Science Park 104, 1098 XG Amsterdam, The Netherlands}
	
	\date{\today}
	
	\begin{abstract}
		Floppy modes {--- deformations that cost zero energy ---} are central to the mechanics of a wide class of systems.
		For disordered systems, such as random networks and particle packings,
			it is well-understood how the number of floppy modes is controlled by the topology of the connections.
			Here we uncover that symmetric geometries, present in e.g. mechanical metamaterials, can feature
			an unlimited number of excess floppy modes that are absent in generic geometries, and in addition can support floppy modes that are multi-branched.	
			We study the number $\Delta$ of excess floppy modes by comparing generic and symmetric geometries with identical topologies, and show that $\Delta$ is extensive, peaks at intermediate connection densities, and exhibits mean field scaling. We then develop an approximate yet accurate cluster counting algorithm that captures these findings. Finally, we leverage our insights to design metamaterials with multiple folding mechanisms.
	\end{abstract}
	
	%\pacs{62.20.-x,62.20.D-,63.50. Lm, 64.60.ah}
	
	\maketitle
	
	Floppy modes (FMs) play a fundamental role in the mechanics of a wide variety of disordered physical systems, from elastic networks \cite{pebblegame1,pebblegame2,alexander,wouterz,wouterPRL,siddesign1,siddesign2} to jammed particle packings  \cite{Jam2003,mvhreview,Sidreview}. Floppy modes also play a role in many engineering problems, ranging from robotics to deployable structures, where the goal is to design structures that feature one or more {\em mechanisms} \cite{Paulino}. Mechanisms  are collections of rigid elements linked by flexible hinges, designed to allow for a collective, floppy motion of the elements.  More recently, floppy modes and mechanisms have received renewed attention in the context of mechanical metamaterials, which are architected materials designed to exhibit anomalous mechanical properties, including negative response parameters, shape morphing, and self-folding \cite{lakes1987,milton92,grima2000auxetic, lakes2001extreme, mullin2007pattern, bertoldi2008mechanics, buckley, siddesign1,siddesign2,cube,cocochain,hierarchy,MMreview17}. An important design  strategy for mechanical metamaterials borrows the geometric design of mechanisms, and replaces their  hinges by flexible parts which connect stiffer elements \cite{MMreview17}.
		In all these examples, understanding how the geometric design
		controls the number and character of the floppy modes plays a central role.
	
{	For systems consisting of objects with a total of $n_d$ degrees of freedom, connected by hinges that provide $n_c$ constraints,
		the number of {non-trivial} floppy modes $n_f$ and states of self-stress $n_{ss}$ are related by Maxwell-Calladine counting as {$n_f-n_{ss} = n_d-n_c -n_{rb}$ where $n_{rb}$ counts the trivial rigid body modes ($n_{rb}=3$ in two dimensions)}  \cite{MC}.}
		For generic, disordered systems $n_f$ and $n_{ss}$ can  be determined separately
		from the connection topology \cite{pebblegame1,pebblegame2}, but when symmetries are present such approaches break down and counting only yields the difference $n_f-n_{ss}$. For example, spring lattices which feature perfectly aligned bonds can generate  {\em excess floppy modes} (and associated states of self stress) that disappear under generic perturbations and thus escape topology-based counting methods \cite{pebblegame1,pebblegame2,align,guestlattice,guest2,woutersquare,lubensky1,lubensky2}.
		Mechanical metamaterials often feature symmetric architectures where excess floppy modes (EFMs) may arise, but their geometries are more complex than spring lattices \cite{MMreview17,scott,bertoldi2010negative,overvelde2012compaction,shim2013harnessing,bertoldi2017harnessing,arvind}.

	\begin{figure}[!t]
		\includegraphics[width=1\linewidth]{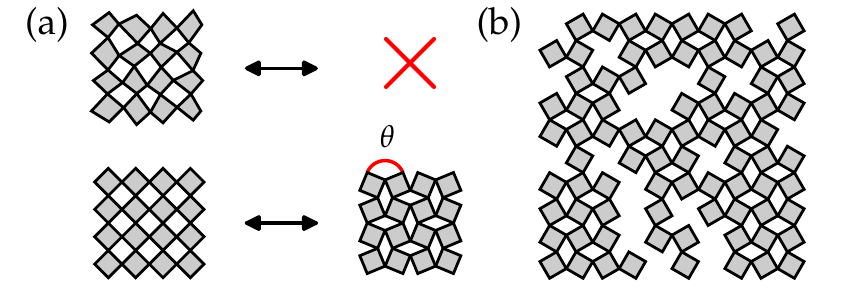}
		\caption{(color online) (a)  $N \times N $ systems of generic quadrilaterals are rigid for $N\ge 3$, {but have a floppy, ''hinging'' mode characterized by the opening angle  $\theta\in\left[0,\pi\right]$ for perfectly symmetric squares.} (b) Diluted square tiling {$(N=10, \rho =0.8)$}.
		}
		\label{fig:1}
	\end{figure}

		We focus on understanding the EFMs of a geometry which underlies a range of metamaterials \cite{grima2000auxetic,mullin2007pattern, bertoldi2008mechanics,cocochain,hierarchy,MMreview17,mexicanhat,dias}: rigid quadrilaterals connected by flexible hinges.
		%We systematically compare the generic case of disordered quadrilaterals and the symmetric case where all elements are squares. In both cases,
{We define $n_s$ and $n_g$ as the number of non-trivial floppy modes
for symmetric systems consisting of squares, and stress-free generic systems obtained by randomly displacing the corners of linked squares with magnitude $\epsilon = 0.1$ \cite{footnote_eps}. } Each quadrilateral has three degrees of freedom (DOF), and in a fully connected lattice each quadrilateral has four connections in the bulk,  and less near the boundary. By counting the degrees of freedom and constraints one finds that $M\times N$ lattices of {\em generic} quadrilaterals are rigid {($n_g=0$)}  when $M \ge 3$ and $N \ge 3$; {however, replacing the generic quadrilaterals by equally-sized {\em squares}, such lattices always exhibit an EFM where the squares can counter-rotate~\cite{grima2000auxetic}, {i.e., $n_s=1$} [Fig.~\ref{fig:1}(a)].} Here we address two key issues. First, what is the multiplicity and statistics of EFMs in diluted lattices as considered recently \cite{hierarchy,mexicanhat,dias} [Fig.~\ref{fig:1}(b)]. {Second, do EFMs in diluted lattices possess anomalous properties, and if so, how can we leverage these to embed new functionalities into metamaterials?}
	
	\begin{figure}[!t]
		\includegraphics[width=1\linewidth]{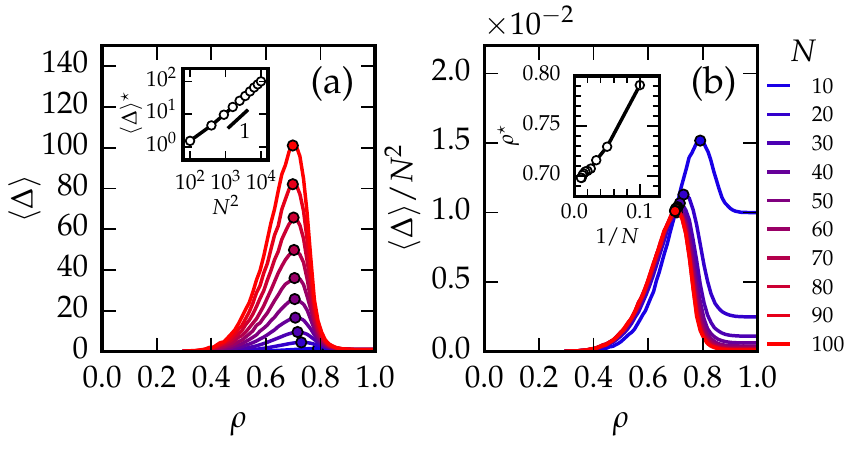}
		\caption{(color online) (a)
			Average number of excess floppy modes as function of the filling fraction $\rho$ for $N$ as indicated. Inset: the peak value
			$\langle\Delta\rangle^\star \approx \beta N^2$ with $\beta \approx 0.01$.
			(b) Scaling collapse of $\langle \Delta \rangle/N^2$ for large $N$. Inset:
			peak position $\rho^\star$ as function of $1/N$ shows clear convergence of $\rho^\star$ to $0.69 \pm 0.01$ for large $N$.}
		\label{fig:2}
	\end{figure}
	
	{\em System and Methods.--- } We consider diluted $N\times N$ lattices of
	quadrilaterals connected by springs of unit stiffness and zero rest length ---
	unless noted otherwise, we use open boundary conditions [Fig.~\ref{fig:1}(b)].
	For each filling fraction $\rho$, we repeatedly remove
		$(1-\rho) N^2$ random quadrilaterals (or links - see below) to obtain a specific connection topology, and for each topology we calculate  $n_s$ and  $n_g$.
	
	{\em Random Dilution.---} We focus on the ensemble averaged number of EFMs, $\langle \Delta \rangle := \langle n_s - n_g \rangle$ as function of $\rho$ [Fig.~\ref{fig:2}(a)] \cite{footnotensng}.
{In the dilute limit (small $\rho$), the system breaks up into isolated quadrilaterals for which $ n_s = n_g $ and $\langle \Delta\rangle = 0 $.} In the undiluted limit ($\rho = 1$) discussed above, there are zero $(n_g  =0)$ floppy modes in the generic case and a single $ (n_s = 1) $ counter rotating EFM in the symmetric case, so that $\langle \Delta\rangle = 1$. Strikingly, we find that for intermediate densities, $\langle \Delta \rangle$ is not monotonic but exhibits a maximum $\langle \Delta\rangle^* \gg 1$, unambiguously evidencing the emergence of multiple EFMs in diluted, symmetric systems. To see how $\langle \Delta\rangle$ can become larger than one, consider $m$ disconnected, fully filled $3\times3$ clusters: as each symmetric cluster has an internal hinging EFM absent for generic clusters, $\Delta=m$ for such a hypothetical system.
	While more intricate topologies arise for random dilution,  we will show that the internal hinging motion of clusters of squares are central to EFMs.
	
	{\em Scaling.--- } We find that the number of EFMs, $\langle\Delta\rangle$, follows mean-field scaling. First, we find that its
	maximum, $\langle\Delta\rangle^*$, grows linearly with $N^2$, which implies
	that the peak density of EFMs, $\beta:=\langle\Delta\rangle^*/N^2$, is a constant [inset Fig.~\ref{fig:2}(a)].
	Second, we can collapse our data as
	$\langle\Delta\rangle/ N^2 = f\left(\rho\right)$, with
	the peak location approaching a constant as $\rho^\star = \rho_0 +  \alpha/N$ with $\rho_0\approx0.69$, $\alpha \approx 1.00$ [Fig.~\ref{fig:2}(b)]. Third, we found that the distribution of $\Delta$ at fixed $\rho$ is Gaussian (not shown). Finally, we have
	also studied random {bond removal, i.e. removal of individual links}, and find a very similar scaling collapse and peak location.
	All this strongly suggests that EFMs occur with a constant density and finite correlation length (see S.I. for details).

	\begin{figure}[!t]
		\includegraphics[width=1\linewidth]{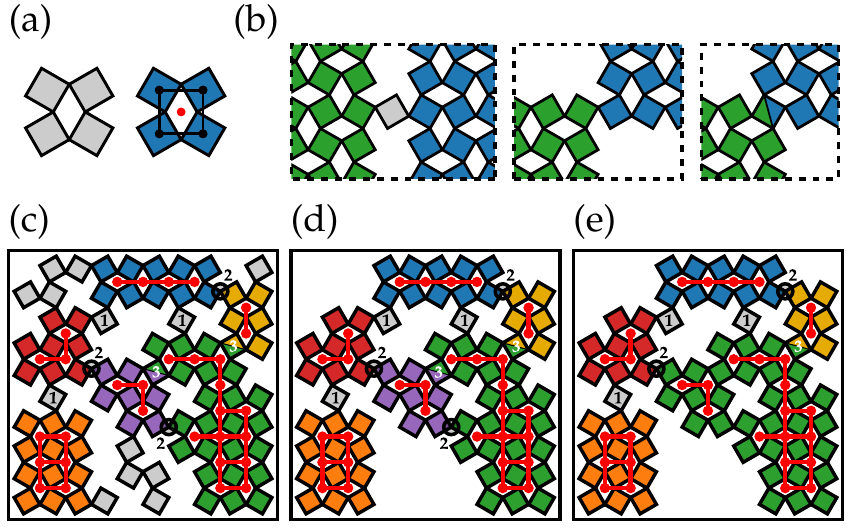}
		\caption{(color online) (a) A 4-block is formed by four connected quadrilaterals, and can be identified by filling a (red) dot on the dual grid. (b) Type-1, type-2 and type-3 connectors. (c) Clusters are adjacent 4-blocks,  and feature straight connections between dots on the dual grid.  All quadrilaterals
			that belong to a single cluster share a color; quadrilaterals belong to zero (grey),
			one (color) or two clusters (bi-color). Grey quadrilaterals that are connected via two hinges at
			clusters are type-1 connectors; all other grey quadrilaterals
			are ''remaining quadrilaterals'', which are either isolated,
			connected to other grey quadrilaterals only, or connected at only one side to a cluster.
			(d) Clusters in pruned system where all remaining quadrilaterals are removed. (e) Cluster merging; note that merged clusters may contain 4-blocks that are not adjacent. %\la{[rode lijnen = dual grid, moet ergens genoemd worden]}
		}
		\label{fig:3}
	\end{figure}

	{\em Clusters and Counting. ---}
{The number of floppy modes in generic systems, $n_g$, can exactly be determined by the pebble-game \cite{pebblegame1,pebblegame2}, and therefore to study $\Delta$ one can study $n_s$ and vice versa.
		We now construct a counting argument for the number of floppy modes in the symmetric systems $(n_s)$.} In this argument we interpret diluted geometries as systems of $N_c$ clusters and connectors. We define {\em clusters} as groups of adjacent 4-blocks, where 4-blocks are groups of four quadrilaterals connected in a loop [Fig.~\ref{fig:3}(a)]. Each cluster has a single internal hinging mode in addition to their three translational/rotational floppy modes. Clusters can be connected by three types of connectors [see Fig.~\ref{fig:3}(b)]: a type-1 connection introduces one constraint and consists of a single quadrilateral; a type-2 connection introduces two constraints and occurs when two adjacent clusters touch; a type-3 connection  introduces three constraints and occurs when two clusters share a quadrilateral.
	
	Generally,
		a number of loosely connected {\em remaining quadrilaterals} are neither connector nor part of any cluster [Fig.~\ref{fig:3}(c)].
		We remove these quadrilaterals and focus on {\em pruned} systems that solely consist of clusters and connectors [Fig.~\ref{fig:3}(d)], and refer to the quantities in the pruned systems by accents (e.g., $\Delta'=n_s'-n_g'$) --- later we will present evidence that $\Delta'$ and $\Delta$ are very close.
	
{ In the simplest counting arguments one ignores the states of self stress, leading to
an estimate $\Delta_0 = n'_s-n'_g$. We now consider that each cluster has three global and one internal
degree of freedom, and estimate $n'_s$ via the difference of the number of degrees of freedom associated with the clusters, $4 N_c$, and the number of constraints between clusters, $\Sigma_{ij}C_{ij}$, where the connection number
	$C_{ij}:= n_{1,ij}+2 n_{2,ij}+3 n_{3,ij}$ counts the number of constraints between clusters $i$ and $j$, and $n_{k,ij}$ denotes the number of type-$k$ connectors. Noting that we need to subtract the three rigid body motions, this yields $n'_s = 4 N_c-\Sigma C_{ij}/2-3$, and
$\Delta_0:= 4 N_c-\Sigma C_{ij}/2-3-n'_g$.  However, we will show that
		$\Delta_0$ is quite different from $\Delta'$ obtained numerically, as symmetries lead to degeneracies between connectors.}

	%\la{Overgang naar volgende alinea niet duidelijk. Ik zou hier een overgang/tegenstelling creeren door duielijk te maken dat het naive counting argument er ver naast zit vanwege degenerate constraints.}
	
	Degenerate constraints arise when $C_{ij} > 4$. To see how such degeneracies arise, consider a system of two  clusters. Without connections these have a total of eight DOFs, but even if $C_{ij}$ is very large, this system must feature at least three rotational/translational and one global hinging mode, so that for $C_{ij} > 4$, constraints must be redundant (see S.I.). To take these degeneracies into account, we define a sequence of increasingly accurate predictions $\Delta_0, \Delta_1,\dots, \Delta_{\infty}$ by progressively
		eliminating  degeneracies. To define $\Delta_1$,  we merge all pairs of clusters for which $C_{ij} \ge4$, and then for the resulting set of clusters and connectivities define $N_{1,c}, C_{1,ij}$, and define $\Delta_1:= 4N_{1,c}-\Sigma C_{1,ij}/2-3-n'_g$.
		As cluster merging may yield new pairs of heavily connected clusters, we iterate this procedure
		to obtain $\Delta_2,\Delta_3,\dots$. Eventually, no cluster merging is possible
		when all $ C^n_{ij} \le 4 $ [Fig.~\ref{fig:3}(e)], and the final result for $\Delta_n$ is denoted as $\Delta_{\infty}$.
	
	\begin{figure}[!t]
		\includegraphics[width=1\linewidth]{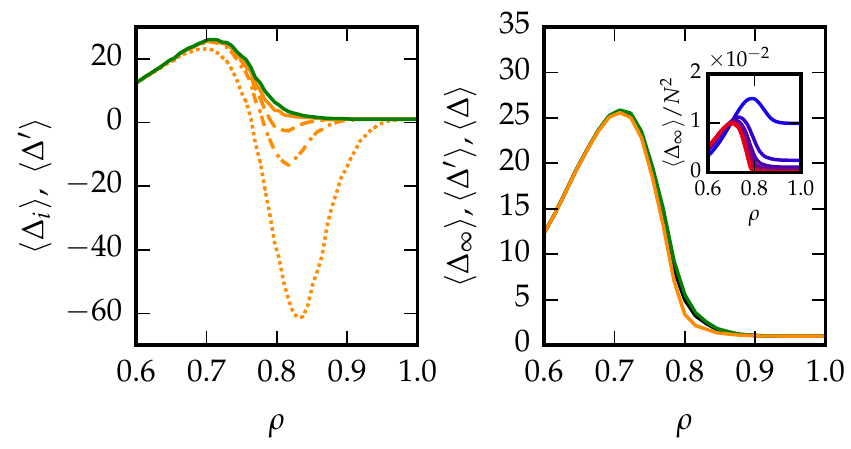}
		\caption{(color online) (a)~ Counting predictions $\langle \Delta_0 \rangle, \langle \Delta_1 \rangle, \langle \Delta_2 \rangle$ (orange dotted, dash-dotted, dashed) and $\langle \Delta_{\infty} \rangle$ {(solid orange)} compared to the exact result $\langle \Delta' \rangle$ {(green)} for $N=50$.  (b) Comparison of $\langle \Delta_{\infty} \rangle$ (orange), $\langle \Delta^\prime \rangle$ (green) and $\langle \Delta \rangle$ (black) for $N=50$. Inset: asymptotic scaling collapse of $\langle \Delta_\infty \rangle$ for $N=10,20,\dots,90$ (blue to red).}
		\label{fig:4}
	\end{figure}
	
	{\em Results.---}
	We now compare the results $\Delta_i$ obtained by our iterative cluster counting argument to the numerical results for $\Delta^\prime$ and $\Delta$ [Fig.~4].
		We find that while $\Delta_0$ captures the peak in the number of EFMs, it
		fails in the high density regimes where degeneracies are abundant. Cluster merging significantly improves the results, and $\Delta_{\infty}$ is found to be within a few percent of $\Delta'$ [Fig.~\ref{fig:4}(a)].
		Moreover, we find that pruning has a minor effect, as $\Delta_{\infty}$, $\Delta'$ and $\Delta$ are very close [Fig.~\ref{fig:4}(b)]. Finally,
		$\Delta_{\infty}$ displays a scaling collapse closely matching that of $\Delta$ [inset Fig.~\ref{fig:4}(b)]. We conclude that our approach of cluster merging and counting is able to accurately capture the numerically observed data for the number of EFMs, thus providing fresh insight into how floppy modes and states of self stress proliferate in complex geometries with symmetries.
	
	%Examples of rare cases that cause deviations between
	%$\Delta_\infty$ and $\Delta^\prime$ are discussed in the S.I.
	% \la{[en hoe zit het met $\Delta^\prime$ dan. Deze is niet exact omdat we niet alle degeneracies kunnen elimineren. Dit moeten/kunnen we benoemen als we ook verschillen met $\Delta_{\infty}$ benoemen]}
	
	\begin{figure*}[!t]
		\includegraphics[width=1\linewidth]{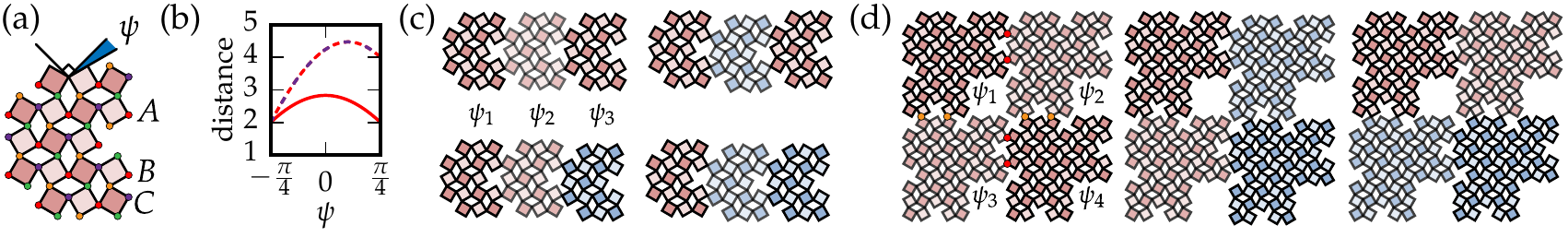}
		\caption{(color online)
			(a) A cluster with positive opening angle
			$\psi:= (\theta-\pi/2)/2$. The colored dots indicate the four types of hinges.
			(b)
			The distance $AB$ (full)  is symmetric in $\psi$ as
			$A$ and $B$ correspond to the same hinge type ($ AB = 2\sqrt{2}\cos(\psi)$)
			while  $AC$ (dashed) is not invariant under $\psi \leftrightarrow -\psi$, as
			$A$ and $C$ correspond to different hinge types ($AC = AB + 2\sin(\psi + \pi/4)
			$). (c) Clusters connected by type-2 connectors at points of the same hinge type can
			either have equal or opposite $\psi$ (red: $\psi>0$, blue: $\psi<0$). Fixing
			$\psi_1>0$ to break the global inversion symmetry $\psi_i \leftrightarrow -\psi$, the four branches
			of motion are characterized by the freely assignable signs of $\psi_2$ and $\psi_3$.
			(d) Compound cluster structure consisting of four identical unit cells that exhibits
			three branches of motion - see text and S.I.
		}
		\label{fig:5}
	\end{figure*}

	{\em Multi-branch planar folding mechanisms.---}
	{Shape-morphing metamaterials can act as deployable structures, deforming along a well defined path under external actuation \cite{milton1995,pellegrino97,Paulino,cube,hierarchy,MMreview17,scott,cocochain,oribox,orimaha}.}
		Their geometric design often follows from a mechanism with a single floppy mode.
		To embed multiple shape changes in a metamaterial, rather than designing structures with
		multiple floppy modes which lead to a continuous family of shapes,
		mode-competition and frustration \cite{hierarchy,kagomemultimode,KatiaNature},
		a better route may be to consider structures with a single floppy mode but multiple, {\em discrete} branches. Multi-branch mechanisms occur in, e.g., Origami \cite{scott,arvind}, but we are not aware of any 2D mechanisms that allow multi-branch behavior without self-intersections. Here we show that  by leveraging the symmetries in systems of hinging squares
		on  can design multi-branched mechanisms, consisting of connected clusters whose {\em magnitude} of hinging is coupled, while the {\em signs} of the hinging motion  of the clusters can only take on a limited number of values.
	
	To understand the design of multi-branched mechanisms, we first consider the hinging motion of a single cluster.  We define the opening angle $\psi:=(\theta-\pi/2)/2$, color the squares according to their alternating rotating motion, and note that we can distinguish four link types, associated with connections at the north, east, south or west tip of a dark square [Fig.~\ref{fig:5}(a)]. The crucial observation is that the distance between two links
	is invariant under opening angle inversion ($\psi \leftrightarrow -\psi$) if and only if both links are of equal type [Fig.~\ref{fig:5}(a)]. Specifically, if we consider links that are vertically aligned, such as $A$, $B$ and $C$ in Fig.~\ref{fig:5}(a), we find that $AB$ is {\em symmetric} in $\psi$, while $AC$ is not [Fig.~\ref{fig:5}(b)].
	
	We obtain multi-branched mechanisms by connecting clusters by type-2 connectors located at hinges $A$ and $B$, whose distance is symmetric in the opening angle. We define
	$\psi_1, \psi_2, \dots$ such that for a global hinging mode all opening angles are equal.
	The symmetry of $AB(\psi)$ then implies that, e.g., the clusters 1 and 2 can be in a ''homogeneous'' state where $\psi_1=\psi_2$, and in second state, where $\psi_1=-\psi_2$ \cite{footnote_signsym}.
	Clusters that can be connected by two type-2 connectors can easily be constructed,
	and strips of $m$ of such coupled clusters form a compound structure with one continuous hinging degree of freedom, and precisely $2^{m-1}$ discrete branches, where $\{\psi_i\} =\{\psi_1,(-1)^{s_2}\psi_1\, \dots,(-1)^{s_m}\psi_1\}$;  here the binary variables  $\{s_2,\dots,s_{m}\}$ characterize the different branches [Fig.~\ref{fig:5}(c)].
	
	More complex situations where the number of branches is different from a power of two arise when clusters are connected in more complex topologies.
	We designed a cluster which features the single and double-bump edge structures along both its vertical and horizontal edges, and connect these into a $2\times 2$ lattice [Fig.~\ref{fig:5}(d)]. As the (horizontal) connections between cluster 1 and 2, respectively 3 and 4, are of a different link type than the (vertical) connections between cluster 1 and 3 respectively 2 and 4, the sign of $\psi_i$ has to be consistent along either rows or columns.
	Fixing $\psi_1>0$, we obtain two consistent row configurations  with $\{ s_i\} = \{0,0,0,0\}$ and $\{0,0,1,1\}$; similarly, consistent column configurations
	are $\{ s_i\} = \{0,0,0,0\}$ and $\{0,1,0,1\}$.
	It is easy to show that these are all allowed possibilities, and as we double count the homogeneous configuration, this example constitutes a
	{\em three-branch} mechanism. We note that the  table of allowed branches $s_2, s_3$ is equivalent to the truth table of a NAND gate.
	Extending this design to $m\times n$ tilings, we obtain
	$2^{m-1}$ different column arrangements, $2^{n-1}$ different row arrangements yielding
	$2^{m-1} + 2^{n-1}-1$ branches. For small $m\le 3$ and $n\le 3$ this already allows to make mechanisms with 1, 2, 3, 4, 5, 7, 8, 9, 11 and 15 branches. This suggests that our design approach, where multiple blocks with symmetric motions are connected together, and where loop-like compatibility conditions determine which branches can be realized, constitutes a powerful new method to design complex, multi-branched mechanisms.
	
	{\em Outlook and Discussion.---} We have characterized the excess floppy modes that arise in
		diluted lattices of square elements, and have introduced an approximate yet accurate methodology to identify and count such excess modes. As our method is developed and tested for square elements, it would be interesting to see if it is similarly successful for other systems with symmetries, including diluted triangular and kagome lattices. Furthermore, we used our insights to design planar mechanisms with multiple discrete folding motions. Such mechanisms, which feature a single continuous degree of freedom but multiple discrete branches provide new design avenues for (soft) robotics, deployable structures and mechanical metamaterials \cite{rus2015design,Paulino,overvelde2016three,MMreview17}, as well as opening up systematic strategies for	 the design of multistable structures \cite{scott}. Finally, open questions include how one designs compounds with any integer number of branches, or with arbitrary ''truth'' tables for the hinging signs of the clusters  $\{ s_i\}$. %\la{OK!}

	\begin{acknowledgments}		
		We acknowledge discussion with W. Ellenbroek,  B. G.-g. Chen and S. Guest, and
		funding from the Netherlands Organization for Scientific Research through a VICI grant No. NWO-680-47-609.
	\end{acknowledgments}

	%\bibliographystyle{apsrev4-1}
	%\bibliography{EFMs}

\begin{thebibliography}{48}
\bibitem{footnote_eps}{This perturbation is sufficiently large so that in generic system the eigenvalues of EFMs become large and numerically clearly separated from those of FMs (see S.I.).}
\bibitem{footnotensng}{We note that $n_s$ and $n_g$ are almost two orders of magnitude larger than $\Delta$, and that moreover, $n_s$ and $n_g$ have their peaks at significantly lower densities than $\Delta$ (see S.I.).}
\bibitem{footnote_signsym}{Configurations $\{\psi_i\}$ and $\{-\psi_i\}$ are related by a global sign inversion symmetry and are considered to be on the same branch.}
\end{thebibliography}
	
%
	
\end{document}